\documentclass[aps,preprintnumbers,superscriptaddress,twocolumn]{revtex4-1}
\usepackage{graphicx}
\usepackage{bm}
\usepackage{amsmath,amssymb,amsfonts,textcomp}
\usepackage{multirow}
\usepackage{gensymb}
\usepackage{color}
\usepackage{array}
\usepackage{hhline}
\usepackage[dvipsnames]{xcolor}
\usepackage[normalem]{ulem}
\usepackage{verbatim}   
\usepackage{subfigure}  
\usepackage{braket}
\usepackage{float}
\usepackage{subfigure}
\definecolor{fuchsia}{rgb}{0.81,0.38,0.76}
\definecolor{darkgreen}{rgb}{0.36,0.59,0.24}
\definecolor{blue}{rgb}{0.38,0.87,0.79}
\definecolor{orange}{rgb}{0.91,0.61,0.20}
\definecolor{blue}{RGB}{0,0,153} 
\definecolor{blueref}{RGB}{0,0,153} 
\usepackage[colorlinks=true,     linkcolor = blueref,
anchorcolor = blueref,
citecolor = blueref,
filecolor = blueref,
urlcolor = blueref]{hyperref}
\hypersetup{pdfborder=0 0 0}

\begin{document}

\title{Dimer Physics in the Frustrated Cairo Pentagonal Antiferromagnet Bi$_2$Fe$_4$O$_9$}
\author{K. Beauvois} 
 \altaffiliation[]{current address: Institut Laue Langevin, 38000 Grenoble, France}
 \email[]{beauvois@ill.fr}
   \affiliation{Universit\'e Grenoble Alpes, CEA, IRIG, MEM, MDN, 38000 Grenoble, France}
   \affiliation{Institut N\'eel, CNRS \& Universit\'e Grenoble Alpes, 38000 Grenoble, France}
\author{V. Simonet}
 \email[]{virginie.simonet@neel.cnrs.fr}
   \affiliation{Institut N\'eel, CNRS \& Univ. Grenoble Alpes, 38000 Grenoble, France}
\author{S. Petit}
 \affiliation{Laboratoire L\'eon Brillouin, CEA-CNRS, Universit\'e Paris-Saclay, CE-Saclay, 91191 Gif sur Yvette, France}
 \author{J. Robert}
   \affiliation{Institut N\'eel, CNRS \& Univ. Grenoble Alpes, 38000 Grenoble, France}
\author{F. Bourdarot}
   \affiliation{Univ. Grenoble Alpes, CEA, IRIG, MEM, MDN, 38000 Grenoble, France}
\author{M. Gospodinov}
   \affiliation{Institute of Solid State Physics, Bulgarian Academy of Sciences, 1184 Sofia, Bulgaria}
   \author{A. A. Mukhin}
 \affiliation{Prokhorov General Physics Institute, Russian Academy of Sciences, 119991 Moscow, Russia}
 \author{R. Ballou}
   \affiliation{Institut N\'eel, CNRS \& Univ. Grenoble Alpes, 38000 Grenoble, France}
 \author{V. Skumryev}
 \affiliation{Departament de F\'{i}sica,  Universitat Aut\`{o}noma de Barcelona, 08193 Bellaterra, Barcelona, Spain}
  \affiliation{Instituci\'{o} Catalana de Recerca i Estudis Avan\c cats, 08010 Bellaterra, Barcelona, Spain}
\author{E. Ressouche} 
\affiliation{Univ. Grenoble Alpes, CEA, IRIG, MEM, MDN, 38000 Grenoble, France}
\date{\today}

\begin{abstract}

The research field of magnetic frustration is dominated by triangle-based lattices but exotic phenomena can also be observed in pentagonal networks. A peculiar noncollinear magnetic order is indeed known to be stabilized in Bi$_2$Fe$_4$O$_9$ materializing a Cairo pentagonal lattice. We present the spin wave excitations in the magnetically ordered state, obtained by inelastic neutron scattering. They reveal an unconventional excited state related to local precession of pairs of spins. The magnetic excitations are then modeled to determine the superexchange interactions for which the frustration is indeed at the origin of the spin arrangement. This analysis unveils a hierarchy in the interactions, leading to a paramagnetic state (close to the N\'eel temperature) constituted of strongly coupled dimers separated by much less correlated spins. This produces two types of response to an applied magnetic field associated with the two nonequivalent Fe sites, as observed in the magnetization distributions obtained using polarized neutrons. 

\end{abstract}

\maketitle

Magnetic frustration, expected to occur when all spin pair interactions cannot be simultaneously satisfied, is one of the major ingredients at the origin of the flurry of discoveries in magnetic studies for the last 20 years. One of the basic experimental signatures of frustration is the difficulty of a system to order magnetically in spite of significant magnetic interactions, the extreme case being the absence of magnetic order at zero temperature as a consequence of the macroscopic degeneracy of the ground state \cite{Ballou,Moessner,Balents,Moessner2}. In materials where the magnetic moments order eventually at finite temperature, this leads in the region above the ordering temperature but well below the temperature characterizing the strength of the interactions, to a classical spin liquid state also called cooperative paramagnetic, where the magnetic moments are highly correlated although fluctuating. This disordered state can sustain not only well-defined excitations \cite{Robertb} but also zero energy modes, which are the signature of local motions connecting the ground-state spin configurations. These modes acquire a gap on entering the ordered phase \cite{Matan} and are alternatively described in a molecular approach \cite{Tomiyasu}. This rich physics has been well established for triangle-based lattices, for instance the Heisenberg kagome antiferromagnet with nearest-neighbor interactions \cite{Robertb,Zhitomirsky}. Many other exotic manifestations of magnetic frustration have been revealed in quantum systems or with additional ingredients such as strongly anisotropic Hamiltonians \cite{Bramwell,Savary,Kitaev}.

Another direction has been opened with the identification of an equivalent for the pentagonal Cairo lattice in the real material Bi$_2$Fe$_4$O$_9$ \cite{Ressouche}. The Cairo lattice is not based on triangles but on edge-sharing pentagons. It is then still prone to magnetic frustration due to the odd number of bonds in the elementary pentagonal units. This pentagonal lattice has a complex connectivity with three- and fourfold connected sites at variance with triangle-based lattices, which fosters alternative ways to accommodate frustration. This leads in Bi$_2$Fe$_4$O$_9$ to an unconventional ground state consisting of an orthogonal arrangement of the magnetic moments. This classical ground state was also obtained theoretically, as well as other interesting phases, including, in the presence of quantum fluctuations, a resonating valence bond liquid or an orthogonal dimer ground state (valence bond crystal) \cite{Raman,Rousochatzakis}. The latter recalls the exactly solvable dimer ground state of the Shastry-Sutherland lattice \cite{ShastrySutherland} largely investigated for its exotic physics \cite{Chung,Shastry}. These findings stimulated further theoretical \cite{Ralko,Urumov,Rojas,Nakano,Isoda,Karlova,Rodrigues} and experimental \cite{Abakumov,Tsirlin,Chattopadhyay,cumby} studies on pentagon-based physics, even spreading beyond the field of magnetism. In spite of this interest, an experimental determination of the Hamiltonian of the prototypical material Bi$_2$Fe$_4$O$_9$ has never been reported yet that would solely ascertain the crucial influence of frustration on its exotic properties.  

In this Letter, we present the experimental determination of the magnetic interactions in Bi$_2$Fe$_4$O$_9$ materializing a Cairo lattice with spins 5/2 using inelastic neutron scattering. A minimum of five exchange interactions allows us to account for the magnetic order and for the associated excitations including a peculiar quasiflat mode. We also show the magnetic density maps measured using polarized neutron scattering under a magnetic field. They reveal a correlated state above the ordering temperature, resulting from the hierarchy of interactions and characteristics of the underlying dimer physics. 

\begin{figure}[h!]
\includegraphics[width=0.75\columnwidth]{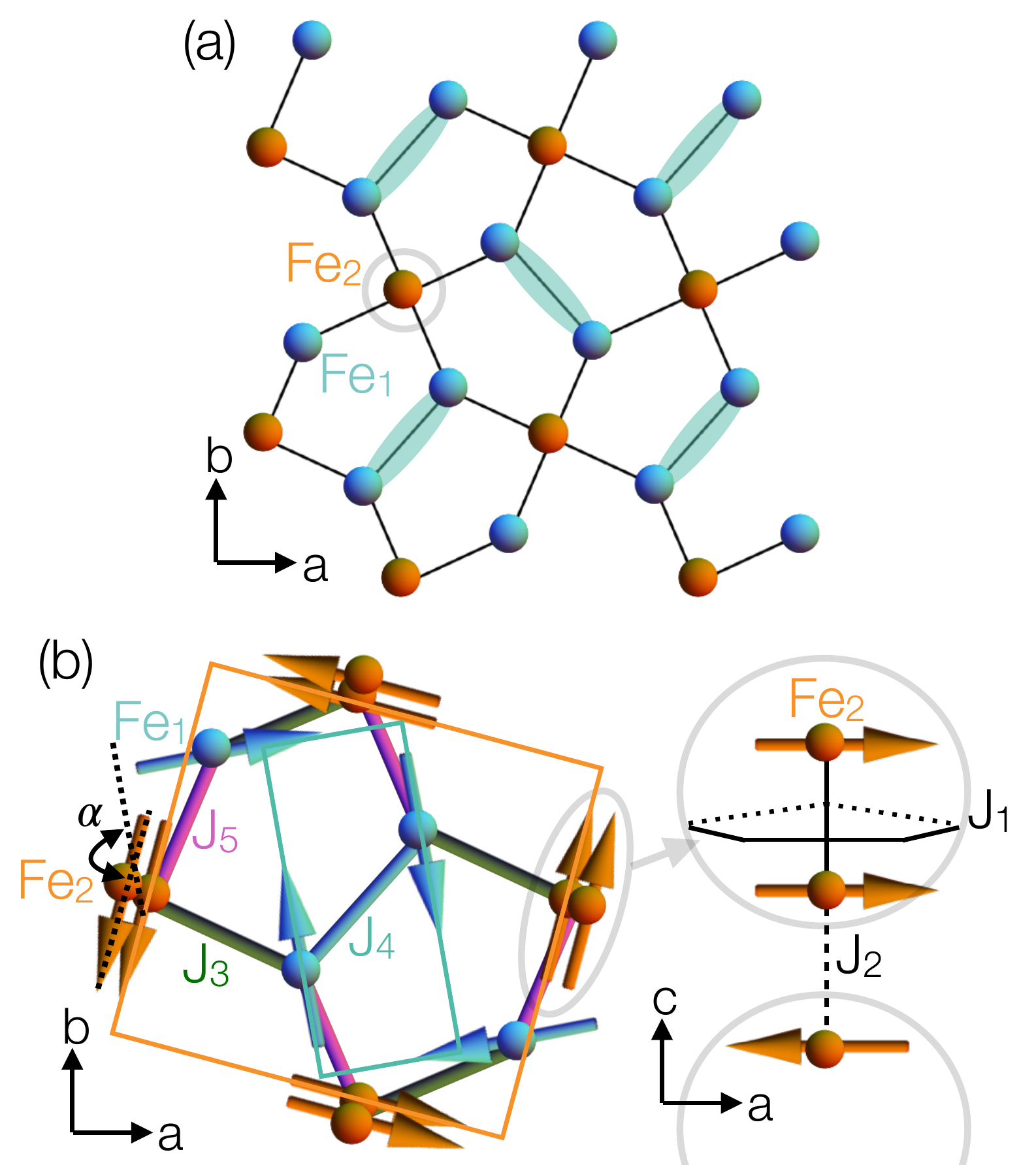}
\caption{(a) $ab$ plane projection of the Fe atoms in Bi$_2$Fe$_4$O$_9$ forming an equivalent of the Cairo pentagonal lattice. The Fe$_1$ (in blue) and Fe$_2$ (in orange) have a different connectivity. Blue ellipses underline strongly coupled antiferromagnetic Fe$_1$ spins. (b) Magnetic arrangement stabilized below $T_N$: the orange and blue rectangles materialize the two sets of orthogonal antiferromagnetic pairs in the $ab$ plane \cite{Ressouche}. The Fe$_2$ form ferromagnetic pairs sandwiching the Fe$_1$ planes. The five exchange interactions within and between the pentagonal planes are labeled.
}
\label{FiglatticeExchange}
\end{figure}

The unit cell of the orthorhombic oxide Bi$_2$Fe$_4$O$_9$ contains eight magnetic Fe$^{3+}$ ions equally distributed on two different Wyckoff sites of the $Pbam$ space group: 4$h$ for Fe$_1$ and 4$f$ for Fe$_2$. These sites have a different connectivity and different oxygen coordination, tetrahedral for Fe$_1$ and octahedral for Fe$_2$. They form a lattice closely related to the Cairo pentagonal one with noticeable differences (see Fig. \ref{FiglatticeExchange}): the site with fourfold connectivity in the perfect lattice is actually constituted by a pair of Fe$_2$ atoms located below and above the pentagonal plane. The projected lattice in the $ab$ plane is also slightly distorted compared to the perfect Cairo lattice as far as the bond lengths and bond angles are concerned. The Heisenberg Hamiltonian with isotropic exchange interactions is a good starting point to describe the magnetic properties of this frustrated lattice of Fe$^{3+}$ ions ($J=S=5/2$, $L=0$)

\begin{equation}
\mathcal{H}=\sum_{\langle i,j\rangle} J_{ij}\mathbf S_i\cdot\mathbf{S}\mathversion{normal}_j
\label{eq1}
\end{equation}
where $\mathbf {S}_i$ is a spin operator and $J_{ij}=J_1$-$J_5$ are the five superexchange interactions between pairs of spins, inferred from the structure \cite{Ressouche}. There are three exchange couplings in the \textit{ab} plane ($J_3$, $J_4$, and $J_5$) and two additional ones out of the plane of the pentagons ($J_1$ and $J_2$), which connect only the stacked Fe$_2$ [see Fig. \ref{FiglatticeExchange} (b)]. Bi$_2$Fe$_4$O$_9$ exhibits below $T_{ N}\sim240$\,K a long-range antiferromagnetic order characterized by a propagation vector $\vec{k}$ = (1/2, 1/2, 1/2). The resulting spin configuration was elucidated by neutron diffraction (see Fig. \ref{FiglatticeExchange}) and is made of two sets of orthogonal pairs of antiferromagnetic spins in the $ab$ plane corresponding to each Fe site, with a global rotation by an angle $\alpha=155^{\circ}$ relative to each other \cite{Ressouche}. Despite the deviation of the experimental system from  the perfect Cairo lattice, this orthogonal magnetic structure matches the one identified in theoretical studies on the perfect lattice and is shown to be rather robust while varying the ratio of the exchange interactions and increasing the quantum fluctuations \cite{Rousochatzakis}. This peculiar magnetic structure actually results from both frustration and complex connectivity. 

All our experiments were performed on a single crystal of Bi$_2$Fe$_4$O$_9$ of dimensions $\sim2.5\times2\times1.5$\,mm$^3$, grown by the high flux temperature solution method using a flux of Bi$_2$O$_3$. The magnetization distributions were obtained from two neutron scattering experiments performed on the CRG D23 two-axis diffractometer at Institut Laue Langevin (ILL) in its polarized neutron mode, with an incoming neutron wavelength $\lambda= 2.37$ \AA\ from a graphite-Heusler double monochromator configuration. Measurements were performed with a magnetic field of 6 T applied along the sample $\bf{c}$ axis at $T= 250$\,K (paramagnetic state) and along the $\bf{a}$-$\bf{b}$ crystallographic direction at $T= 250$\,K and 15 K (ordered state) (see Supplemental Material \cite{supmat}). Inelastic neutron scattering (INS) experiments were performed on the CRG IN22 triple-axis spectrometer at ILL in an orange cryostat at a constant final wave vector $k_f=2.662$ \AA$^{-1}$, with an energy resolution equal to about 1\,meV. The sample was oriented in order to access the ($h$, $h$, $\ell$) scattering plane.

\begin{figure}[t]
\includegraphics[width=1\columnwidth]{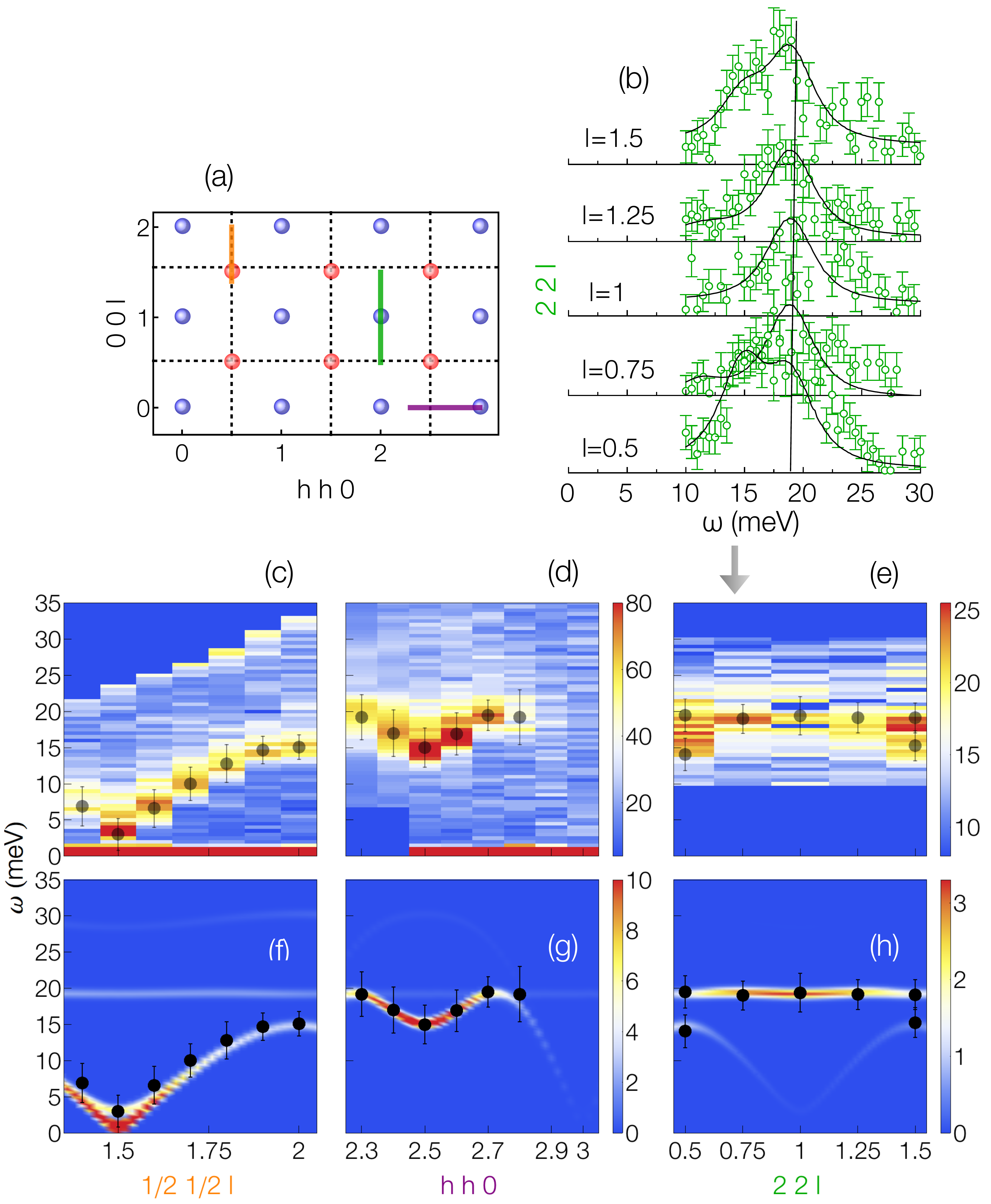}
\caption{(a) Sketch of the scattering plane ($h$, $h$, $\ell$) investigated by INS in Bi$_2$Fe$_4$O$_9$, with the nuclear (blue spheres) and magnetic (red spheres). Bragg peak positions and the measurement cuts along the (1/2, 1/2, $\ell$), ($h$, $h$, 0) and (2, 2, $\ell$) reciprocal space directions. 
(b) Along (2, 2, $\ell$), series of measured constant-{\bf Q} energy scans and fit of the excitations (black lines). Measured (c)-(e) and calculated (f)-(h) dynamic structure factor $S({\bf Q},\omega)$ using the exchange constants of Table \ref{TabExchangeCst} and a small single-ion anisotropy term constraining the spins in the $ab$ plane. The color scale of the calculations was truncated for the spectra along (1/2, 1/2, $\ell$), ($h$, $h$, 0) in order to emphasize the weaker flat mode along (2, 2, $\ell$). The width of the calculated excitations was taken as the energy resolution of 1\,meV. The empty (filled) black points on top of the measured (calculated) $S({\bf Q},\omega)$ give the fitted energy positions of the experimental spin wave dispersion. }
\label{FigSpinWaves}
\end{figure}

In order to measure the magnetic excitations, several constant-{\bf Q} energy scans have been performed at 1.5\,K along the reciprocal space directions (1/2, 1/2, $\ell$), ($h$, $h$, 0) and (2, 2, $\ell$) sketched in Fig. \ref{FigSpinWaves}(a). The excitations along (2, 2, $\ell$) were fitted using the \textsc{takin} software \cite{Takin2016b,Takin2017b}, which includes the spin wave model as input and the instrument resolution [see Fig. \ref{FigSpinWaves}(b)] (see Supplemental Material \cite{supmat}). This treatment was necessary to disentangle two overlapping modes. The excitations in the other two directions were simply fitted by the sum of two Lorentz functions,
\begin{equation}
S({\bf Q},\omega)=bg+A(\omega,T)\left(\frac{1}{(\omega-\epsilon)^2 +\sigma^2}+\frac{1}{(\omega+\epsilon)^2 +\sigma^2}\right)
\label{2lorentz}
\end{equation}
where $bg$ is the background, $A(\omega,T)=[1+n(\omega,T)]Z\omega\sigma/2$, $n(\omega,T)=1/(e^{\omega/k_BT}-1)$ is the Bose factor, $Z$ is the weight of the excitations, $\epsilon$ is their energy and $\sigma$ is the half width at half maximum. The resulting energy position of the excitations is reported as black dots in Figs. \ref{FigSpinWaves}(c)-\ref{FigSpinWaves}(h). All {\bf Q} scans have been combined into the experimental maps presented in Figs. \ref{FigSpinWaves}(c)-\ref{FigSpinWaves}(e). The magnetic nature of the excitations has been checked through the temperature dependence. Well-defined spin waves are observed, as expected for this ordered compound. An acousticlike mode emerging from the antiferromagnetic Bragg position (1/2, 1/2, 1/2) is clearly visible, as well as a high-energy branch along ($h$, $h$, 0). Additionally, an almost nondispersive mode located at the energy of about 19\,meV is observed along (2, 2, $\ell$). Higher energy modes are inferred from Raman spectroscopy with two magnetic excitations identified at the energies 32.2 and 58.5\,meV, out of the energy window of our neutron experiments \cite{Ilievb}. 

Our INS measurements were then compared with spin wave calculations performed using the SpinWave software \cite{Petitb,SWb} based on the linear spin wave theory using the Holstein-Primakoff formalism \cite{holstein}. The starting point was the model hamiltonian of Eq. \ref{eq1} involving five isotropic superexchange interactions (see Fig. \ref{FiglatticeExchange}). Two additional constrains were used to limit the number of refined parameters. First, it was shown in \onlinecite{Ressouche} that the rotation angle $\alpha=$155\,$\degree$ between both iron sublattices is obtained for $J_3/J_5=2.15$. A second relation between the exchange interactions was inferred from the Curie-Weiss temperature $\theta_{CW}\approx-1670$\, K estimated from magnetic susceptibility measurements \cite{Ressouche}. We used the local Weiss molecular field model on the stabilized magnetic structure and the equipartition theorem \cite{White}
\begin{equation}
2\times\frac{3}{2}k_B\theta_{CW}={\bf S}_1\sum_{j}J_{1j}<{\bf S}_j>+{\bf S}_2\sum_{j}J_{2j}<{\bf S_j}>,
\label{Relation3}
\end{equation}
where ${\bf S}_1$ (${\bf S}_2$) is the spin on site Fe$_1$ (Fe$_2$). This allowed us to further reduce the model to three independent parameters, which were systematically varied in the calculations. We checked the capability of each of the sets of parameters to reproduce the measured spin waves, as well as the magnetic structure of Bi$_2$Fe$_4$O$_9$ through a real-space mean-field energy minimization of the spin configuration. Finally, a model Hamiltonian compatible with the experiments was obtained with the values of the exchange constants given in Table \ref{TabExchangeCst}. The calculated spin waves are displayed in the lower panels of Figs. \ref{FigSpinWaves}(f)-\ref{FigSpinWaves}(h) and show a very good agreement with the experimental data. The calculations indicate that the spin waves dispersion extends at higher energies up to 80 meV (see Supplemental Material \cite{supmat}), with, in particular, two zone center modes at the energy positions of the Raman excitations, which further validates our model. 

\begin{table}[!h]
	\centering
	\begin{tabular}{|p{2cm}||p{1cm}|p{1cm}|p{1cm}|p{1cm}|p{1cm}|}
		\hline	$J$ (meV) & $J_1$  & $J_2$ & \textcolor{darkgreen}{$J_3$} & \textcolor{cyan}{$J_4$} & \textcolor{fuchsia}{$J_5$} \\
		\hline	IN22 & 3.7(2) & 1.3(2) & 6.3(2)  & 24.0(8) & 2.9(1) \\
		\hline
	\end{tabular}
	\caption{Values of the antiferromagnetic exchange interactions of Bi$_2$Fe$_4$O$_9$ deduced from the INS measurements. The uncertainties on $J_1$, $J_2$, and $J_3$ are estimated from the standard deviations of the fit of the experimental spin wave dispersion.
	The $J_4$ and $J_5$ uncertainties have been obtained by error propagation: $J_5$ is imposed by the constraint $J_3/J_5$ and $J_4$ by the relation (\ref{Relation3}).} 
	\label{TabExchangeCst}
\end{table}

Our analysis establishes that all five interactions are antiferromagnetic. Because of the geometry of the pentagonal lattice, this implies competition between the $J_3$, $J_4$ and $J_5$ interactions within the $ab$ planes confirming the role of magnetic frustration in the stabilization of the 90$^{\circ}$ magnetic order previously reported \cite{Ressouche}.

The antiferromagnetic interaction $J_1$ between the Fe$_2$ pairs of spins sandwiching the pentagonal planes is actually overcome by an effective ferromagnetic coupling resulting from the interactions of the Fe$_2$ spins with the Fe$_1$ spins in the $ab$ planes via $J_3$ and $J_5$. Moreover, this indirect coupling produces the same local field on both Fe$_2$ spins. This configuration leads to a nearly flat optical mode visible around 19 meV in all directions of reciprocal space and associated with these pairs of Fe$_2$ spins [see Fig \ref{FigSpinWaves}(e)]. Along (1/2, 1/2, $\ell$) and ($h$, $h$, 0), this mode was not visible in our experimental data. However, we investigated in purpose the (2, 2, $\ell$) direction where this mode was strongest in the calculations. We indeed observed it experimentally along this reciprocal space direction. This excited state corresponds to the out-of-phase precession of both Fe$_2$ spins around their local field, which does not influence the neighboring Fe$_1$ spins and thus remains localized. A weak dispersion is actually calculated due to the $J_{2}$ interplane couplings in the $c$ direction, which is beyond the instrumental resolution of the experiment (see Supplemental Material \cite{supmat}).

Also noticeable in Table \ref{TabExchangeCst} is the fact that the $J_4$ interaction is significantly stronger than the other ones, which is compatible with the 180$^{\circ}$ superexchange path through the central oxygen ion according to the Goodenough-Kanamori rules \cite{Goodenough}. This hierarchy of interactions results in a lattice with dominant pairs of antiferromagnetically coupled spins Fe$_1$ on almost orthogonal bonds [see Fig. \ref{FiglatticeExchange}(a)], a picture that is expected to survive above the N\'eel temperature. Note that the exchange interactions deduced from our spin wave analysis are slightly different from those obtained from \textit{ab initio} calculations \cite{Pchelkinab,supmat} in the LSDA$+U$ approximation. However, the positive sign of all exchange couplings (antiferromagnetic) and the overall hierarchy of the interactions are identical in both cases.

\begin{table}[!h]
	\centering
	\begin{tabular}{|p{1.5cm}||p{2cm}|p{2cm}|p{2cm}|}
		\hline	T (K) & $m$ ($\mu_{ B}$) & $m_{\rm Fe_1}$ ($\mu_{ B}$) & $m_{\rm Fe_2}$ ($\mu_{ B}$)  \\
		\hline	250  & 0.220(5)& 0.001(5) & 0.045(5)  \\
		\hline	15  & 0.263(5) & 0.014(8) & 0.041(7)\\ 
		\hline
	\end{tabular}
	\caption{Values of the magnetization measured in Bi$_2$Fe$_4$O$_9$ at $T=250$ and 15\,K and under a magnetic field $\mu_0H=6$\,T: macroscopic magnetization $m$ per unit cell (second column) obtained with an extraction magnetometer with the field direction along ${\bf a}$-${\bf b}$ at 15 K and along ${\bf a}$-${\bf b}$ or ${\bf c}$ at $250$\,K;  magnetization per site $m_{\rm Fe_1}$ and $m_{\rm Fe_2}$ (third and fourth columns) deduced from the flipping ratio neutron experiments. At $T=15$\,K the magnetic density was obtained for ${\bf H}\parallel{\bf a}$-${\bf b}$. At $T=250$\,K, the magnetic density is the average of the values obtained for ${\bf H}\parallel{\bf c}$ and ${\bf H}\parallel{\bf a}$-${\bf b}$ since the same macroscopic magnetization is measured.}
	\label{TabMoments}
\end{table}

\begin{figure}[t]
	\includegraphics[width=0.8\columnwidth]{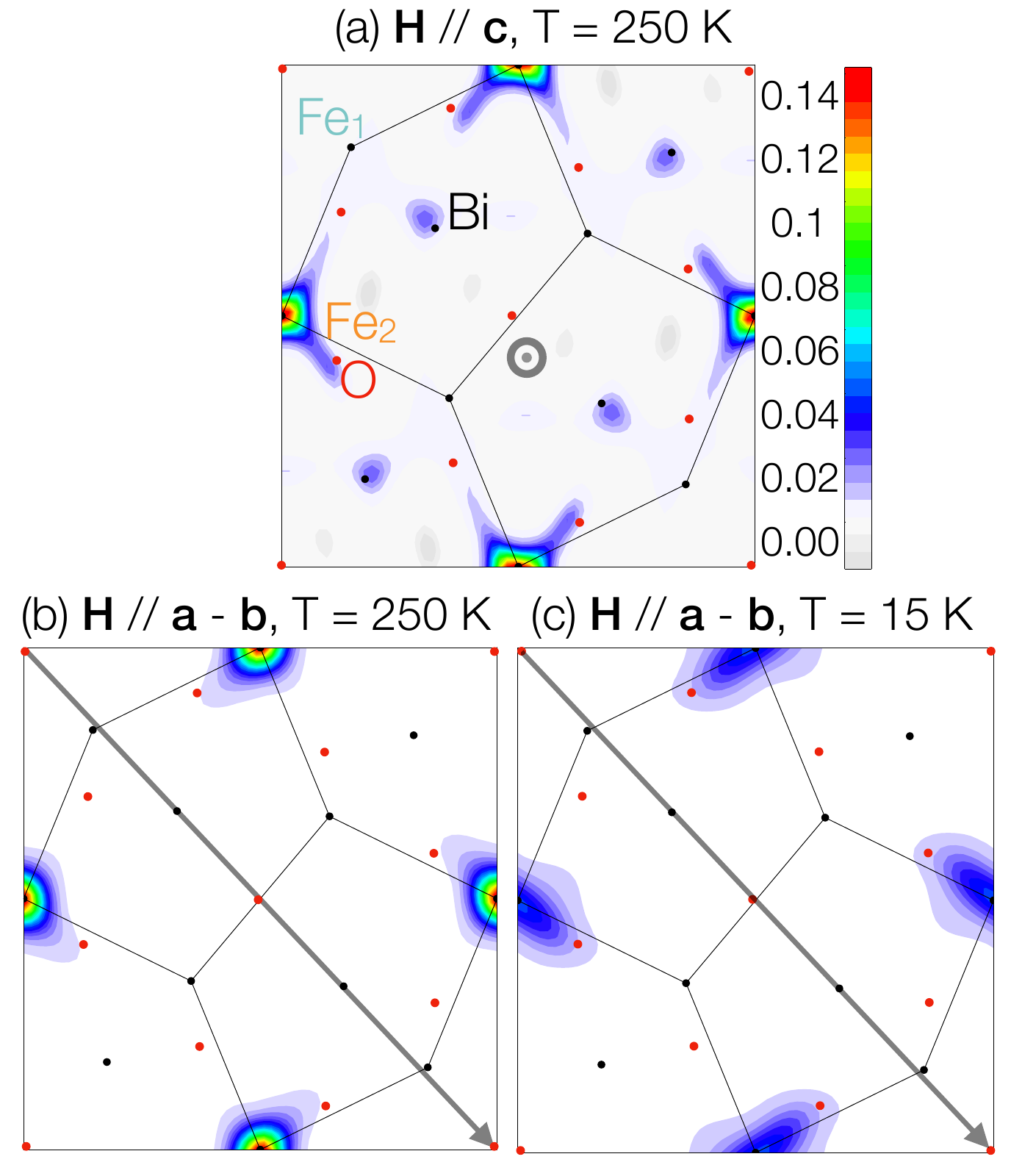}
	\caption{Spatial distributions of the magnetization of Bi$_2$Fe$_4$O$_9$ in $\mu_{ B}$ projected along the $\bf{c}$ axis measured with polarized neutrons under a magnetic field $\mu_0H=6$\,T applied (a) along $\bf{c}$ at $T=250$\,K and (b) along $\bf{a}$-$\bf{b}$ at $T=250$\,K and (c) along $\bf{a}$-$\bf{b}$ at $T=15$\,K (c). }
	\label{FigMagMaps}
\end{figure}

In order to investigate the fingerprint of these Fe$_1$ dimers in the paramagnetic state, we  measured the magnetization distributions under a magnetic field of 6 T with two orientations with respect to the crystallographic axes, as shown in Fig. \ref{FigMagMaps}. The magnetization per site has been extracted from these maps in the dipolar approximation implying a spherical electronic distribution around the atoms. Several fitting processes were performed taking into account the presence of magnetic moments on the iron sites only, or on all the iron, oxygen and bismuth sites. A magnetic contribution can indeed be present on the oxygen atoms and, due to their $6s$ lone electron pairs, on the Bi atoms. The final averaged magnetizations per Fe sites are given in Table \ref{TabMoments}.

Interestingly, in the paramagnetic state [Figs. \ref{FigMagMaps}(a) and \ref{FigMagMaps}(b)], the two iron sites have radically different behaviors: whereas the Fe$_2$ ions carry an induced magnetic moment of 0.045(5) $\mu_{B}$ aligned along the field, the induced magnetization on site Fe$_1$ is vanishingly small (see Table \ref{TabMoments}). This is in contrast with the ordered magnetic moments on both sites refined from previous neutron diffraction experiments below $T_{ N}$ which are rather similar, equal to 3.52 and 3.73 $\mu_{ B}$ respectively \cite{Ressouche}. Our measurements performed with two different directions of the magnetic field yield the same result, which show that the anisotropy is not responsible for this behavior as expected for Fe$^{3+}$ ions with zero orbital angular momentum. The most noticeable difference between the two maps in Figs. \ref{FigMagMaps}(a) and \ref{FigMagMaps}(b) is the presence of magnetic density on the Bi sites for ${\bf H}\parallel{\bf c}$ and not for ${\bf H}\parallel{\bf a-b}$. This could be real or an artifact due to an imperfect reconstruction for ${\bf H}\parallel{\bf c}$ since a smaller number of observations has been used compared to ${\bf H}\parallel{\bf a}$-${\bf b}$. The magnetization distribution has also been measured at low temperature (see Fig. \ref{FigMagMaps}(c)). A larger field-induced polarization on the Fe$_{2}$ compared to the Fe$_{1}$ one is actually preserved in the ordered state. As detailed in Table \ref{TabMoments}, the magnetization on the Fe$_1$ is actually slightly larger at 15\,K than at 250\,K and also more delocalized (see Supplemental Material \cite{supmat}). 

These results point to an original paramagnetic state. A calculation assuming free spins 5/2 in a 6 T field at 250 K yields a field-polarized magnetization of 0.19 $\mu_{ B}$, which is 4 times larger and almost 200 times larger than the one measured on the Fe$_2$ and Fe$_1$ respectively. This suggests, that slightly above $T_{N}$, the Fe$_1$ form an assembly of strongly correlated antiferromagnetic dimers in agreement with the dominant $J_4$, while the Fe$_2$ spins are much less correlated. Since the Fe$_1$ ions form a spin arrangement with great similarities to a Shastry-Sutherland lattice, we suggest that the temperature regime slightly above $T_{ N}$ could be reminiscent of this original physics \cite{ShastrySutherland,Chung,Shastry}. At higher temperature, the correlations among dimers should vanish, while at lower temperature, correlations involving Fe$_2$ spins grow, driving the system to the physics of the Cairo pentagonal lattice. Below $T_{ N}$, the long-range magnetic order is finally triggered by the weakest $J_2$ interaction, connecting the pentagonal planes. Note that Bi$_2$Fe$_4$O$_9$ is not the unique materialization of the Cairo lattice as it is related to a wide family of compounds including the multiferroic $R$Mn$_2$O$_5$ ($R$ a rare-earth/Y, Mn occupying the pentagonal lattice) whose complex magnetodielectric phase diagrams could be investigated in the renewed perspective of pentagonal physics \cite{Radaelli,Chattopadhyay,Peng}.   
  
Our neutron scattering investigation of Bi$_2$Fe$_4$O$_9$ allowed us to achieve a complete determination of its complex magnetic interactions and to unveil various facets of unconventional magnetism including frustration and dimer physics, with distinct behaviors associated with the two inequivalent Fe sites of the pentagonal lattice. The Fe$_1$ ions produce strongly coupled antiferromagnetic pairs of spins dominating the correlated paramagnetic state, whereas in the ordered state, the pairs of Fe$_2$ spins produce original spin dynamics, associated with protected local motions, coexisting with dispersive spin waves. Beyond the canonical examples of frustrated systems like kagome or pyrochlore lattices with first-neighbor interactions, our Letter discloses novel behaviors that should be more generally observed in materials where the frustration is interlocked with complex connectivity and hierarchal interactions.

\acknowledgments We are grateful to J. Debray for the orientation of the sample, the Institut Laue Langevin for providing us with the neutrons, and Navid Qureshi for his help in the data analysis. J. R. thanks B. Canals for the joint development of the software used for the calculations of the spin dynamics presented in the Supplemental Material. Part of this project was supported by Bulgarian National Science Fund BNSF DN -08/9.

\clearpage

\newpage

\onecolumngrid
\begin{center}
	\textbf{\large Supplemental Material: \\
		Dimer Physics in the Frustrated Cairo Pentagonal Antiferromagnet Bi$_2$Fe$_4$O$_9$}\\[.2cm]
	K. Beauvois,$^{1,2,*}$ V. Simonet,$^{2}$, S. Petit,$^3$ J. Robert,$^2$ F. Bourdarot,$^1$ M. Gospodinov,$^4$ A. A. Mukhin,$^5$ R. Ballou,$^2$ V. Skumryev,$^{6,7}$ and E. Ressouche$^1$\\[.1cm]
	{\itshape ${}^1$Universit\'e  Grenoble Alpes, CEA, IRIG, MEM, MDN, 38000 Grenoble, France\\
		${}^2$Institut N\'eel, CNRS \& Universit\'e  Grenoble Alpes, 38000 Grenoble, France\\
		${}^3$Laboratoire L\'eon Brillouin, CEA-CNRS, Universit\'e Paris-Saclay, CE-Saclay, 91191 Gif sur Yvette, France\\
		${}^4$Institute of Solid State Physics, Bulgarian Academy of Sciences, 1184 Sofia, Bulgaria\\
		${}^5$Prokhorov General Physics Institute, Russian Academy of Sciences, 119991 Moscow, Russia\\
		${}^6$Departament de F\'{i}sica,  Universitat Aut\`{o}noma de Barcelona, 08193 Bellaterra, Barcelona, Spain\\
		${}^7$Instituci\'{o} Catalana de Recerca i Estudis Avan\c cats, 08010 Bellaterra, Barcelona, Spain\\}
	${}^*$Current address: Institut Laue Langevin, 38000 Grenoble, France. Electronic address: beauvois@ill.fr\\
	(Dated: \today)\\[1cm]
\end{center}

\setcounter{equation}{0}
\setcounter{figure}{0}
\setcounter{table}{0}
\setcounter{page}{1}
\renewcommand{\theequation}{S\arabic{equation}}
\renewcommand{\thefigure}{S\arabic{figure}}
\renewcommand{\bibnumfmt}[1]{[S#1]}
\renewcommand{\citenumfont}[1]{S#1}

\section{Magnetization maps}

To obtain the magnetic density maps, the flipping ratio method was used. Bragg reflexions were collected up to $\sin\theta/\lambda=0.36$\,\AA$^{-1}$ and the intensities were reduced and corrected for extinction. The polarization for the incident neutron beam was 0.90(1). Accurate magnetic structure factors were extracted from the flipping ratios using the Cambridge Crystallography Subroutine Library \cite{CCSL}. The magnetization distribution in real space was reconstructed using a uniform prior density and the Maximum of Entropy technique \cite{Maxent} in order to minimize the noise and truncation effects. 28 and 49 independent magnetic structure factors $F_M({\bf Q})$ were used in the reconstruction for $\bf{H}\parallel \bf{c}$ and $\bf{H}\parallel \bf{a-b}$ respectively, as well as a value of the macroscopic magnetization measured with an extraction magnetometer at Institut N\'eel. 

\begin{figure}[h!]
	\includegraphics[width=0.8\columnwidth]{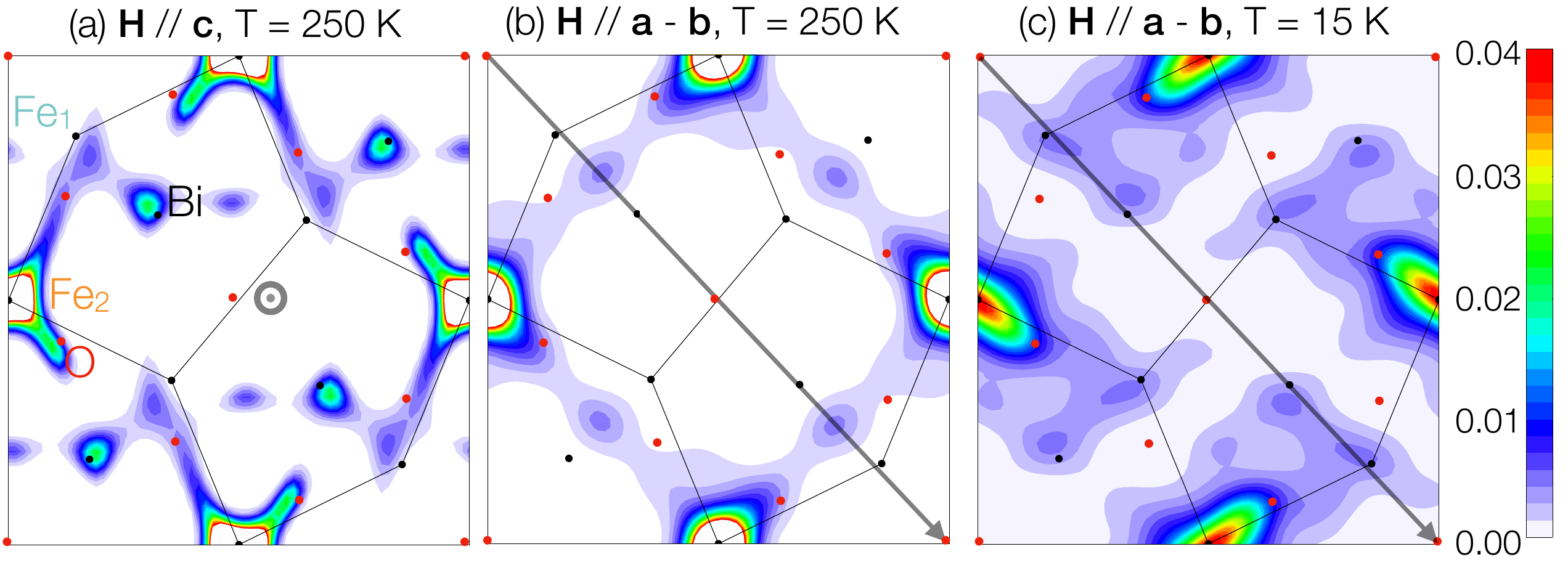}
	\caption{Same spatial distribution density maps of Bi$_2$Fe$_4$O$_9$ as those presented in the main paper (Figure 3) but drawn with a different intensity scale for the magnetization in order to emphasize the weak features. This highlights the delocalization of the magnetization in the ordered state compared to the paramagnetic state.}
	\label{FigMagMaps2}
\end{figure}

\begin{figure}[h!]
	\includegraphics[width=0.6\columnwidth]{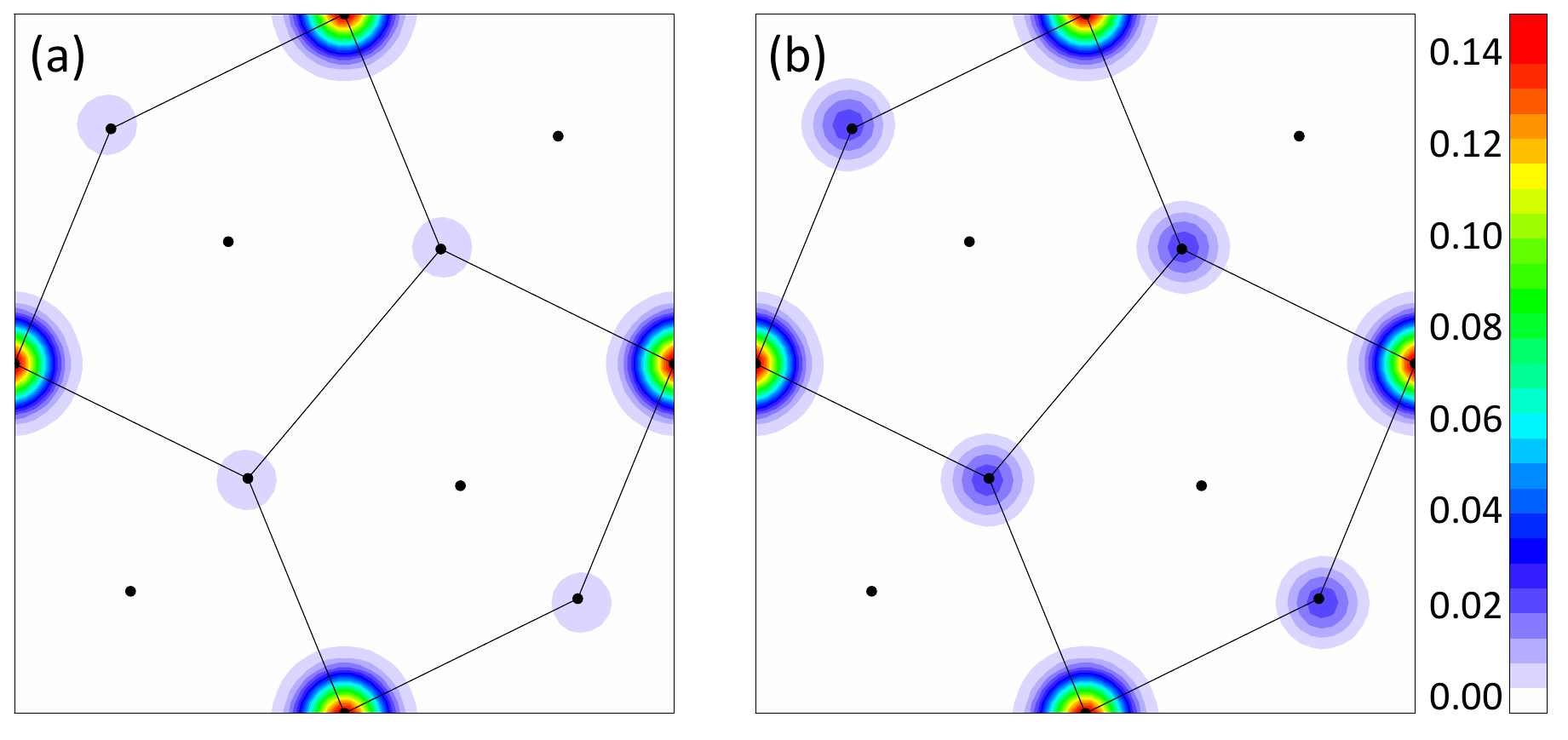}
	\caption{Calculation of the spatial distribution density maps of Bi$_2$Fe$_4$O$_9$ in the dipolar approximation with 0.006 $\mu_{\rm B}$ per Fe$_1$ and 0.049 $\mu_{\rm B}$ per Fe$_2$ (a) or 0.016 $\mu_{\rm B}$ per Fe$_1$ and 0.049 $\mu_{\rm B}$ per Fe$_2$ (b). These maps have the same intensity scale than those presented in Figure 3 of the main paper and give an insight about the sensitivity of the technique.}
	\label{FigMagMaps2}
\end{figure}

\clearpage

\section{Magnetic excitations}

\subsection{Inelastic neutron scattering}

\begin{figure}[h!]
	\includegraphics[width=0.23\columnwidth]{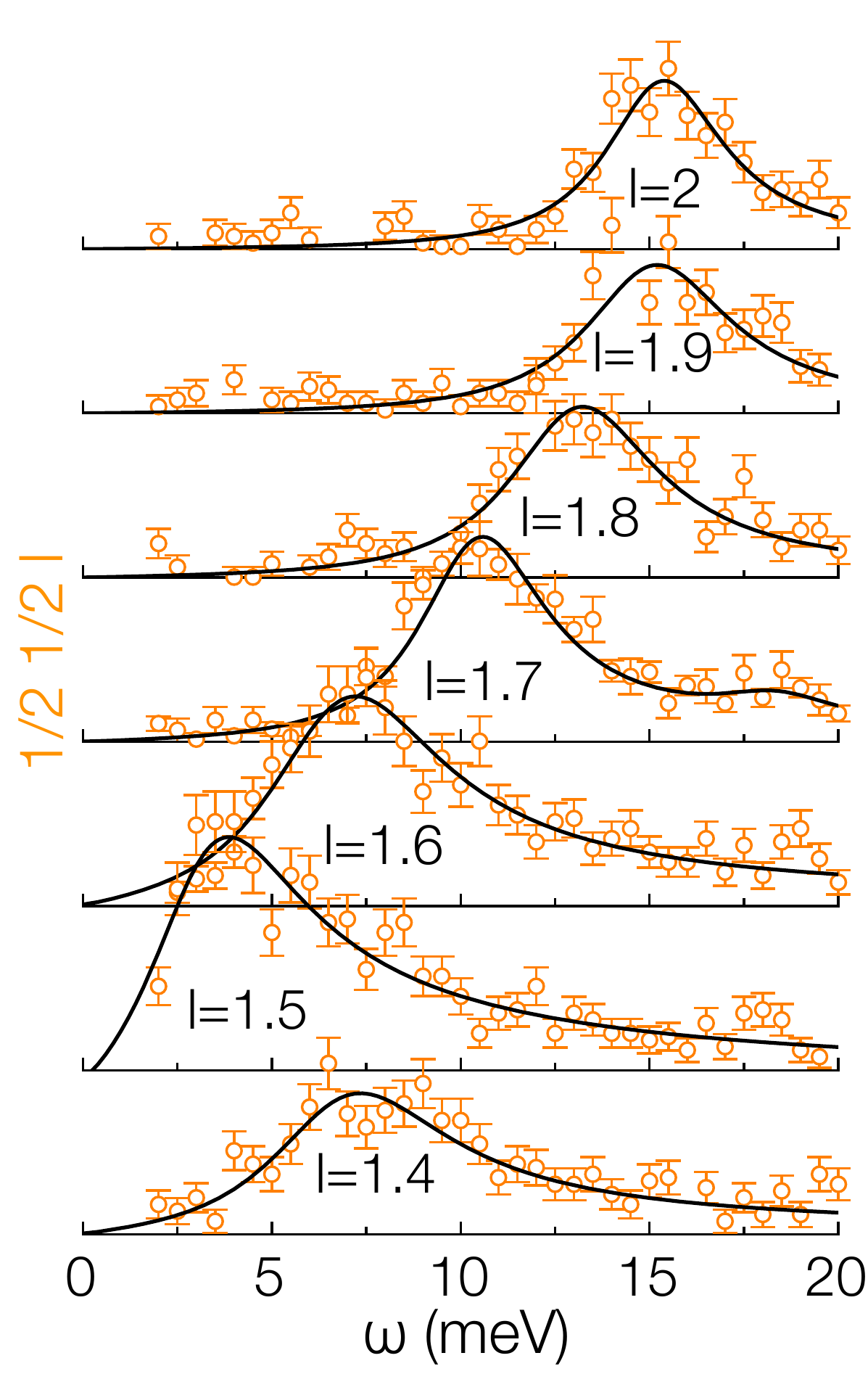}
	\hspace{1cm}
	\includegraphics[width=0.23\columnwidth]{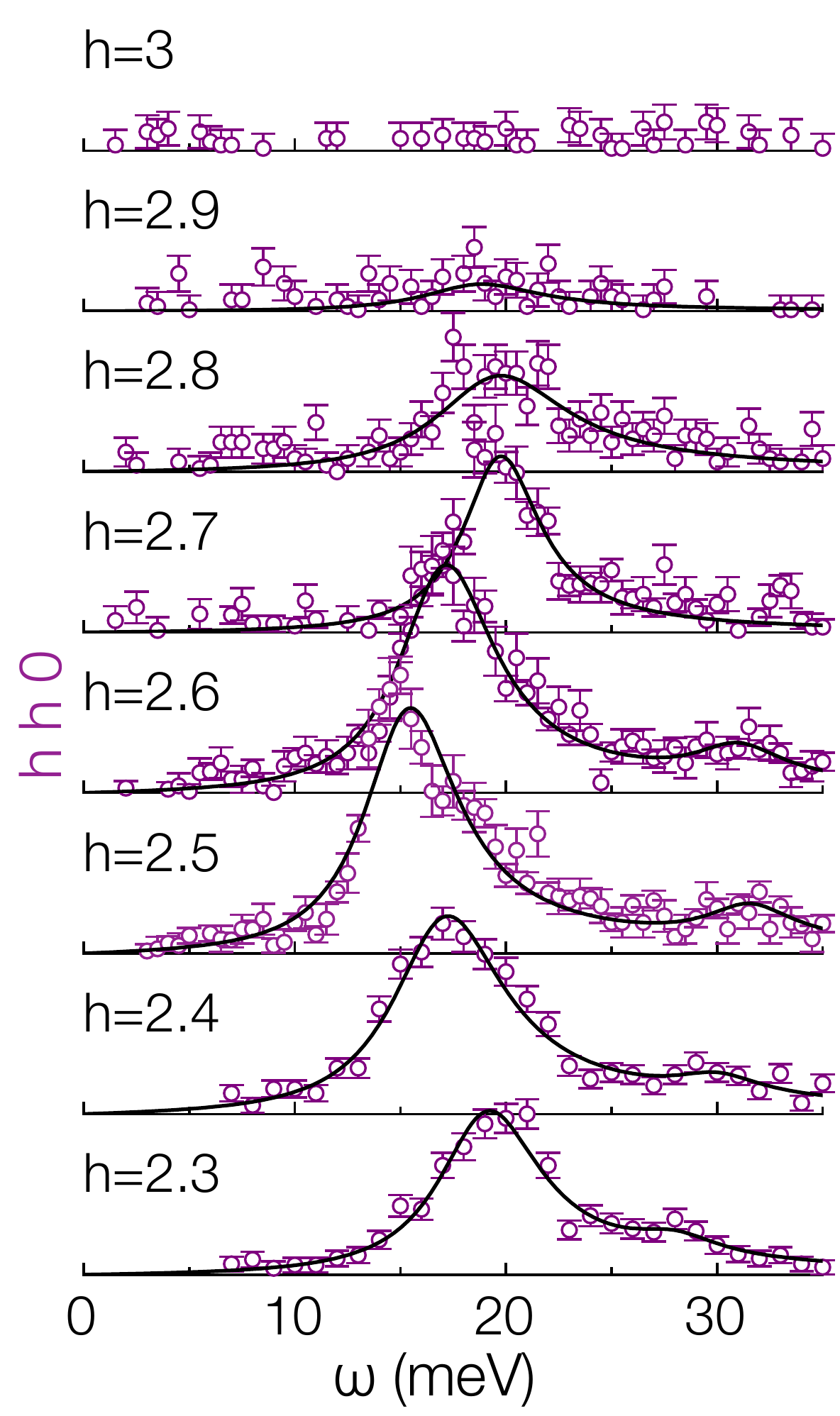}
	\caption{Series of measured constant-$\bf{Q}$ energy scans along (1/2, 1/2, $\ell$) (orange circles) and ($h$, $h$, 0) (purple circles) and fit of the excitations using the model described in the text.}
	\label{FigPlotcuts}
\end{figure}

In order to fit the magnetic excitations along (2, 2, $\ell$), we used the open-source software, Takin, for triple-axis experiments analysis \cite{Takin2016,Takin2017}. We simulated the experimental spectra by convoluting a theoretical dynamical structure factor $S(\bf{Q},\omega)$ model with the instrumental resolution of the IN22 triple axis spectrometer. The resolution function has been calculated using the Popovici algorithm. We used as theoretical model two simple dispersions, one sinusoidal and one flat, for excitations with a lorentzian shape of full width at half maximum equal to 4\,meV. The input of the software were the spin wave dispersions and intensities calculated with the SpinWave program allowing a first fit for each constant-$\bf{Q}$ $\omega$-scan of the global intensity scale factor and of the offset in $\omega$ only. In a second step, all the parameters were let free to vary and a fit of the dispersions, intensities and width of the excitations was performed. This procedure allowed us to distinguish the quasiflat mode from the lower dispersive mode and to determine precisely its energy position at about $20$\,meV. The excitations in the other two directions (1/2, 1/2, $\ell$), ($h$, $h$, 0) were simply fitted by the sum of two Lorentz functions as explained in the main paper.

\clearpage
\subsection{Spin waves calculations}

\begin{figure*}[h!]
	\includegraphics[width=0.6\columnwidth]{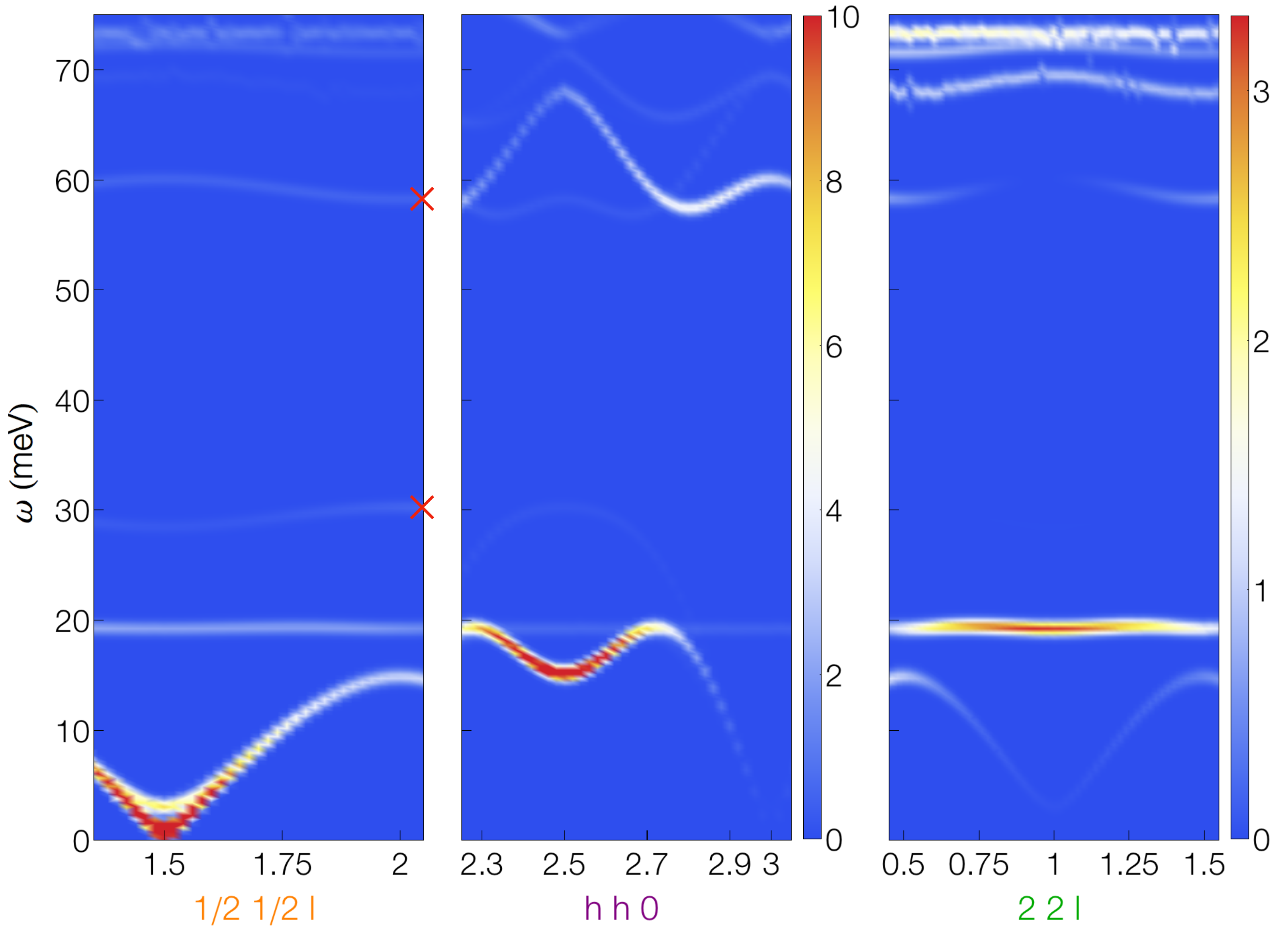}
	\caption{Calculated dynamic structure factor of Bi$_2$Fe$_4$O$_9$ up to 75\,meV  along the reciprocal space directions (1/2, 1/2, $\ell$), ($h$, $h$, 0) and (2, 2, $\ell$) showing the complete magnetic spectrum. This was obtained using the SpinWave software \cite{Petit,SW} with the five exchange interactions reported in Table I of the main article and a single-ion anisotropy within the $ab$-plane of 0.03 meV. The red crosses indicate the two magnetic excitations identified in Raman scattering \cite{Iliev}.}
	\label{FigPlotSimu}
\end{figure*}

\clearpage

\section{Additional information about the quasiflat excitations}

\begin{figure*}[h]
	\includegraphics[width=0.6\columnwidth]{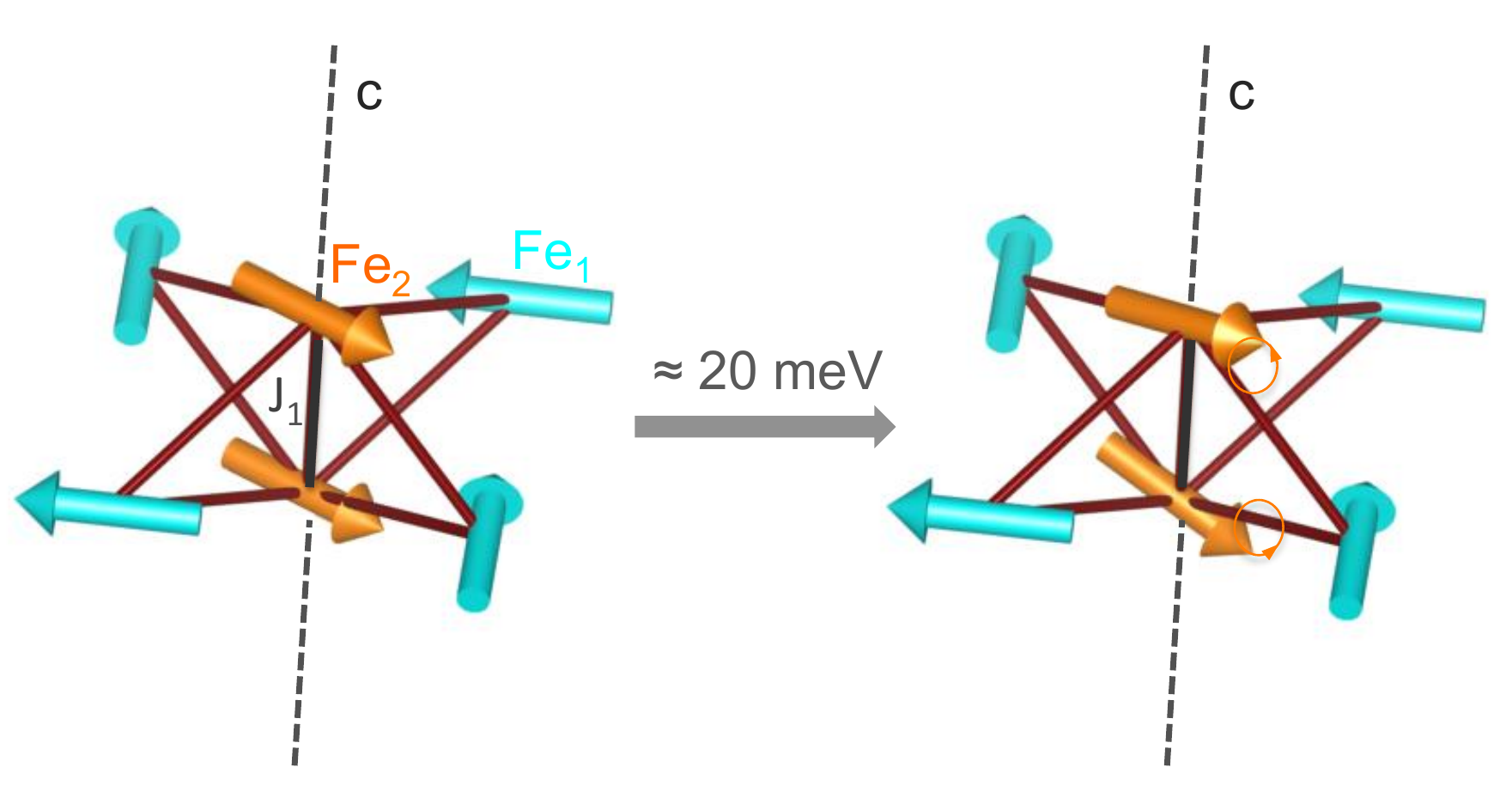}
	\caption{Schematic description of the frustrated mechanism responsible for the quasiflat mode observed at 19 meV in the inelastic spectrum. It involves the out-of phase precession of both Fe$_2$ spins of one ferromagnetic pair connected via a triangular exchange path to four Fe$_1$ neighbors.}
	\label{FigPlotSimu1}
\end{figure*}

As explained in the main text, the spin excitation spectrum includes a peculiar nearly flat optical mode. It involves deviations of the Fe$_2$ spins from their equilibrium position, that remain decoupled from the fluctuations of the four Fe$_1$ to which they are in principle connected in triangular pathways (see Figure \ref{FigPlotSimu1}). For the sake of simplicity, we consider in this reasoning a single pair $({\bf S}_{1i}, {\bf S}_{2i})$ in each unit cell $i$. We assume this pair is located on the $y$ axis, hence the index $i$ refers to the $x$ axis. The case of the second pair (along $x$) can be treated in the same way, the index $i$ being associated to the $y$ axis. Furthermore, we set $J_2 = 0$.\\

For the purpose of this derivation, we define the equilibrium and fluctuations as: ${\bf S}_{1i} = {\bf S}_{1i}^o + {\bf s}_{1i}$. The local field ${\bf H}_i$ experienced by the spins of the same pair is also written: ${\bf H}_i = {\bf H}_i^o + {\bf h}_i$ where ${\bf H}_i^o$ is the equilibrium mean value and ${\bf h}_i$ the corresponding fluctuations. \\

Owing to the magnetic structure, we have: ${\bf S}_{1i}^o={\bf S}_{2i}^o = {\bf S}^o$ (the two Fe$_2$ are parallel) and ${\bf H}_i^o = {\bf H}_{i-1}^o = {\bf H}^o$ (according to the propagation vector, the spins of the Fe$_1$ pairs belonging to neighboring cells and connected to a given Fe$_2$ spin are parallel). Up to first order, the equations of motion write:
\begin{eqnarray*}
	\frac{d{\bf s}_{1i}}{dt} &=&  ({\bf H}_i^o \times {\bf s}_{1i} + {\bf H}_{i-1}^o \times {\bf s}_{1i} + {\bf h}_i \times {\bf S}_{1i}^o + {\bf h}_{i-1} \times {\bf S}_{1i}^o)+ J_1 {\bf S}_{2i} \times {\bf S}_{1i} \\
	\frac{d{\bf s}_{2i}}{dt} &=&  ({\bf H}_i^o \times {\bf s}_{2i} + {\bf H}_{i-1}^o \times {\bf s}_{2i} + h_i \times {\bf S}_{2i}^o + {\bf h}_{i-1} \times {\bf S}_{2i}^o)+ J_1 {\bf S}_{1i} \times {\bf S}_{2i}
\end{eqnarray*}
We then define auxiliary variables:
\begin{eqnarray*}
	{\bf D}_i &=& {\bf S}_{1i}-{\bf S}_{2i} = {\bf s}_{1i}-{\bf s}_{2i} \\
	{\bf S}_i  &=& {\bf S}_{1i}+{\bf S}_{2i} =  2{\bf S}^o + {\bf s}_{1i} + {\bf s}_{2i}
\end{eqnarray*}
hence:
\begin{eqnarray*}
	\frac{d {\bf D}_i}{dt}  &=&\frac{d ({\bf s}_{1i}-{\bf s}_{2i})}{dt} \\
	&=&  ({\bf H}_i^o \times {\bf D}_i + {\bf H}_{i-1}^o \times {\bf D}_i + ({\bf h}_i  + {\bf h}_{i-1}) \times ({\bf S}_{1i}^o - {\bf S}_{2i}^o ) ) + 2J_1 {\bf S}_{2i} \times {\bf S}_{1i} \\
	&=& 2{\bf H}^o \times {\bf D}_i + {\bf 0} + 2J_1 {\bf S}_{2i} \times {\bf S}_{1i} \\
	&=& 2{\bf H}^o \times {\bf D}_i + J_1 {\bf S}_i \times {\bf D}_i\\
	& = & (2 {\bf H}^o + 2J_1 {\bf S}^o ) \times {\bf D}_i
\end{eqnarray*}
This equation shows that ${\bf D}_i$ is decoupled from the ${\bf H}$ and ${\bf S}$ fluctuations and only feels the effective magnetic field $(2{\bf H}^o + 2J_1 {\bf S}^o)$. As a result, the spectrum encompasses an infinite number of resonances, (with no dispersion), whose energy essentially depends on the molecular field $( 2{\bf H}^o + 2J_1 {\bf S}^o )$. Similarly, we can derive the equation of motion for $\vec S_i$, leading to the acoustic counterpart: 
\begin{eqnarray*}
	\frac{d{\bf S}_i}{dt}  &=&  \frac{d({\bf s}_{1i}+{\bf s}_{2i)}}{dt}\\
	&=&({\bf H}_i^o \times {\bf s}_i + {\bf H}_{i-1}^o \times {\bf s}_i + ({\bf h}_i \times  + {\bf h}_{i-1}) \times ({\bf S}_{1i}^o + {\bf S}_{2i}^o ) ) + 2J_1 {\bf S}_{2i} \times {\bf S}_{1i} \\
	&=&2 {\bf H}^o \times {\bf s}_i  +2 ({\bf h}_i  + {\bf h}_{i-1}) \times {\bf S}^o 
\end{eqnarray*}

\begin{figure*}[h]
	\includegraphics[width=0.99\columnwidth]{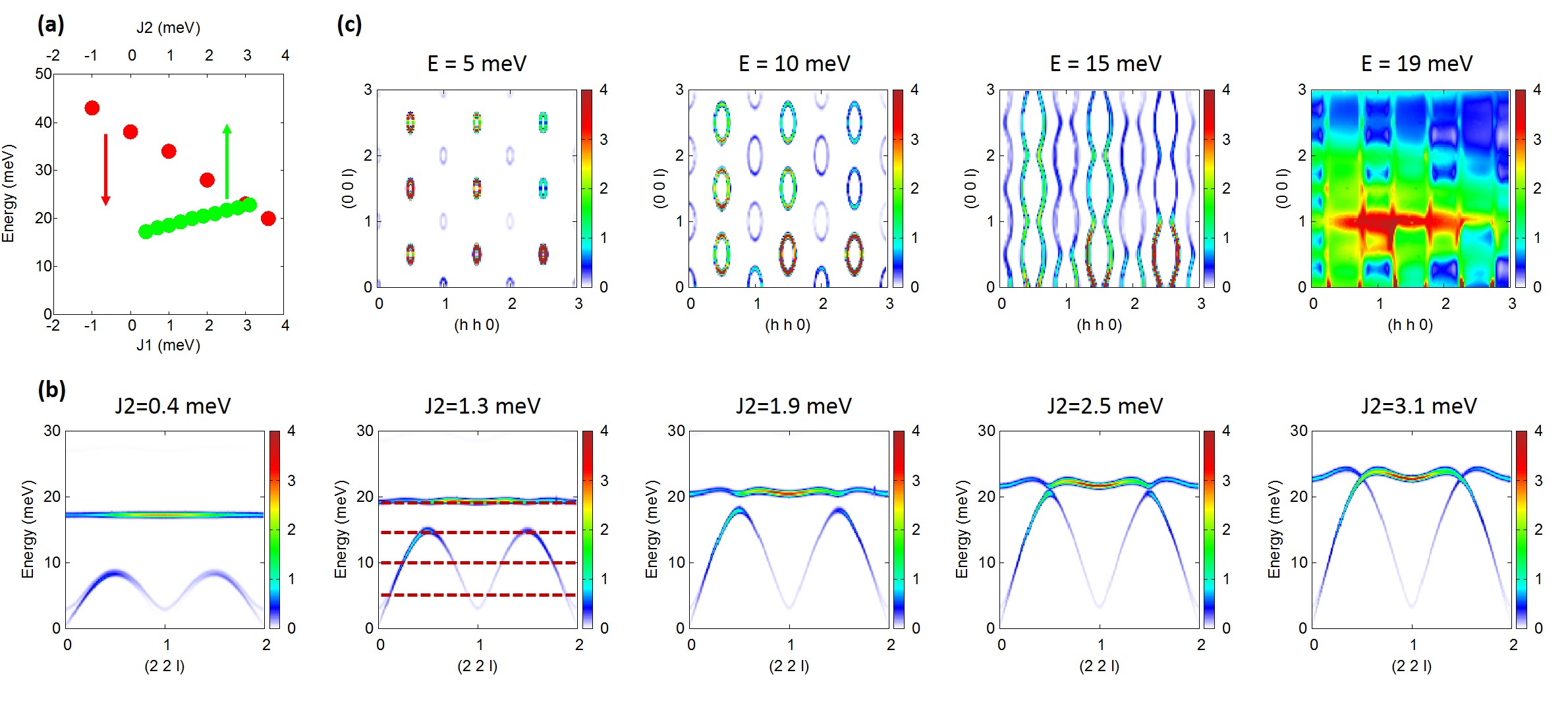}
	\caption{(a): Dependence of the energy position of the quasiflat mode with respect to the value of the $J_1$ (red) and $J_2$ (green) antiferromagnetic interactions connecting directly the two Fe$_2$ spins along the {\bf c} direction. (b) Dispersion of the spinwaves calculated along $(2,2,\ell)$ for several values of $J_2$. The bandwidth of the flat mode is found to increase with increasing $J_2$ up to a point where it connects to the acoustic mode. (c) Cuts at various constant energies across the dispersions for $J_2$=1.3 meV. The energies are shown as dotted lines on the dispersion along $(2,2,\ell)$.}
	\label{FigPlotSimu2}
\end{figure*}

For $J_2 \ne 0$, the calculations performed with the SpinWave software \cite{Petit,SW} show that the dispersion of this mode only depends on $J_2$, connecting the pentagonal planes. For the sake of illustration, Figure \ref{FigPlotSimu2}-b displays the dispersion calculated along $(2,2,\ell)$ for several values of $J_2$ ranging from $J_2=0.1$ up to $J_2=3.1$ meV. Importantly, with increasing $J_2$, the bandwidth of the mode increases and finally connects to the acoustic modes stemming from zone centres. Indeed, the localized character of this mode is ensured by small values of $J_2$. Figure \ref{FigPlotSimu2}-c shows constant energy cuts (at 5, 10, 15 and 19 meV) across the dispersions. Classical cones are observed stemming from the magnetic Bragg peaks, corresponding to the acoustic modes. Those modes become essentially 1-dimensional, forming lines along $\ell$. The flat mode has a peculiar dynamical structure factor, which can be deduced from the 19 meV cut. As expected, it essentially depends on $\ell$, with stronger intensities at $\ell=0,1,2,..$.

\clearpage

\section{Comparison with ab-initio calculations}

\begin{figure*}[h!]
	\includegraphics[width=0.6\columnwidth]{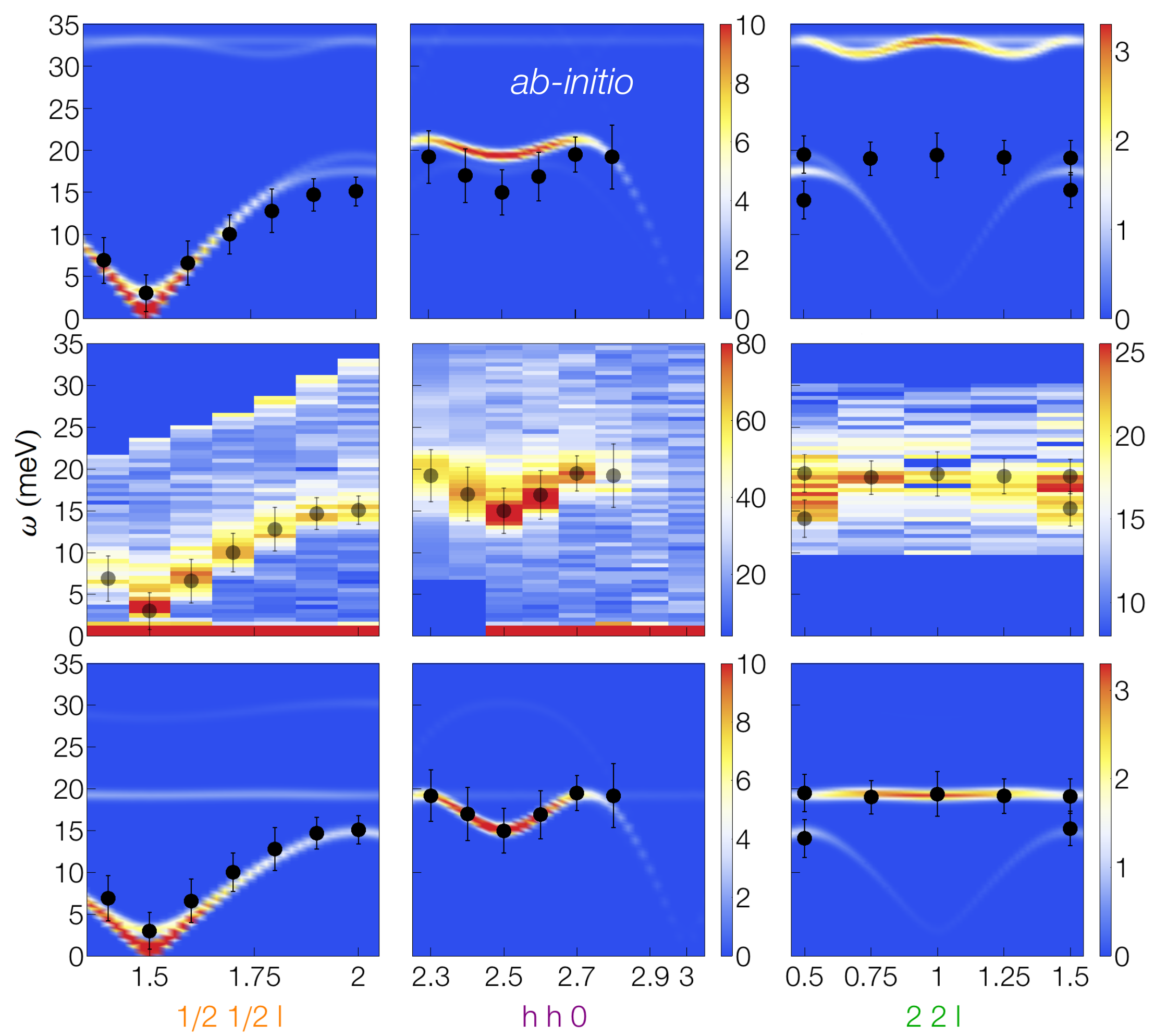}
	\caption{Comparison of the measured spin waves (middle panel) with the ones calculated from the ab initio set of interaction (top panel) and the ones obtained from our spin waves analysis (bottom panel) already shown in Figure 2 of the main article.}
	\label{FigPlotAbinitio}
\end{figure*}

We discuss the difference mentioned in the main text between the exchange interactions deduced from our spin wave analysis of the magnetic spectra measured by inelastic neutron scattering and the ones obtained from ab-initio calculations \cite{Pchelkina}.
In Table \ref{TabExchangeCst}, one can see that the values for the interactions are in the same energy range. There are however some discrepancies and, in particular, our $J_4$ value is twice larger. Moreover, the ab-initio $J_3$/$J_5$ ratio does not yield the correct angle between the Fe$_1$ and Fe$_2$ magnetic sublattices. Figure \ref{FigPlotAbinitio} shows the calculated spin waves from the ab-initio interaction values. Contrary to the spin waves calculated with the set of values deduced from our spin wave analysis, the energy and shape of the dispersing branches do not match well with the experimental results. Due to the lower $J_1$ value, the energy position of the quasiflat mode is too high, at about 30\,meV instead of 19\,meV. 
These discrepancies might be accounted for by some limiting assumptions in the ab-initio calculations that are recalled in the reference \cite{Himmetoglu}.

\begin{table}[!h]
	\centering
	\begin{tabular}{|p{2cm}||p{1cm}|p{1cm}|p{1cm}|p{1cm}|p{1cm}|}
		\hline	$J$ (meV) & $J_1$  & $J_2$ & \textcolor{darkgreen}{$J_3$} & \textcolor{cyan}{$J_4$} & \textcolor{fuchsia}{$J_5$} \\
		\hline	IN22 & 3.7(2) & 1.3(2) & 6.3(2)   & 24.0(8) & 2.9(1) \\
		\hline	Ab-initio & 1.7 & 2.1 & 6.2  & 12.6 & 4.0 \\
		\hline
	\end{tabular}
	\caption{Comparison of the antiferromagnetic exchange interactions of Bi$_2$Fe$_4$O$_9$ deduced from the inelastic measurements and obtained from ab-initio calculations ($U=4.5$, $J_H=1$) \cite{Pchelkina}. The values reported in reference \cite{Pchelkina} for the ab-initio calculations have been multiplied by a factor two to compare with ours due to a different definition of the Hamiltonian.}
	\label{TabExchangeCst}
\end{table}

\clearpage

\section{Classical spin dynamics calculations of the magnetic structure factor}

\begin{figure*}[h!]
	\includegraphics[width=0.9\columnwidth]{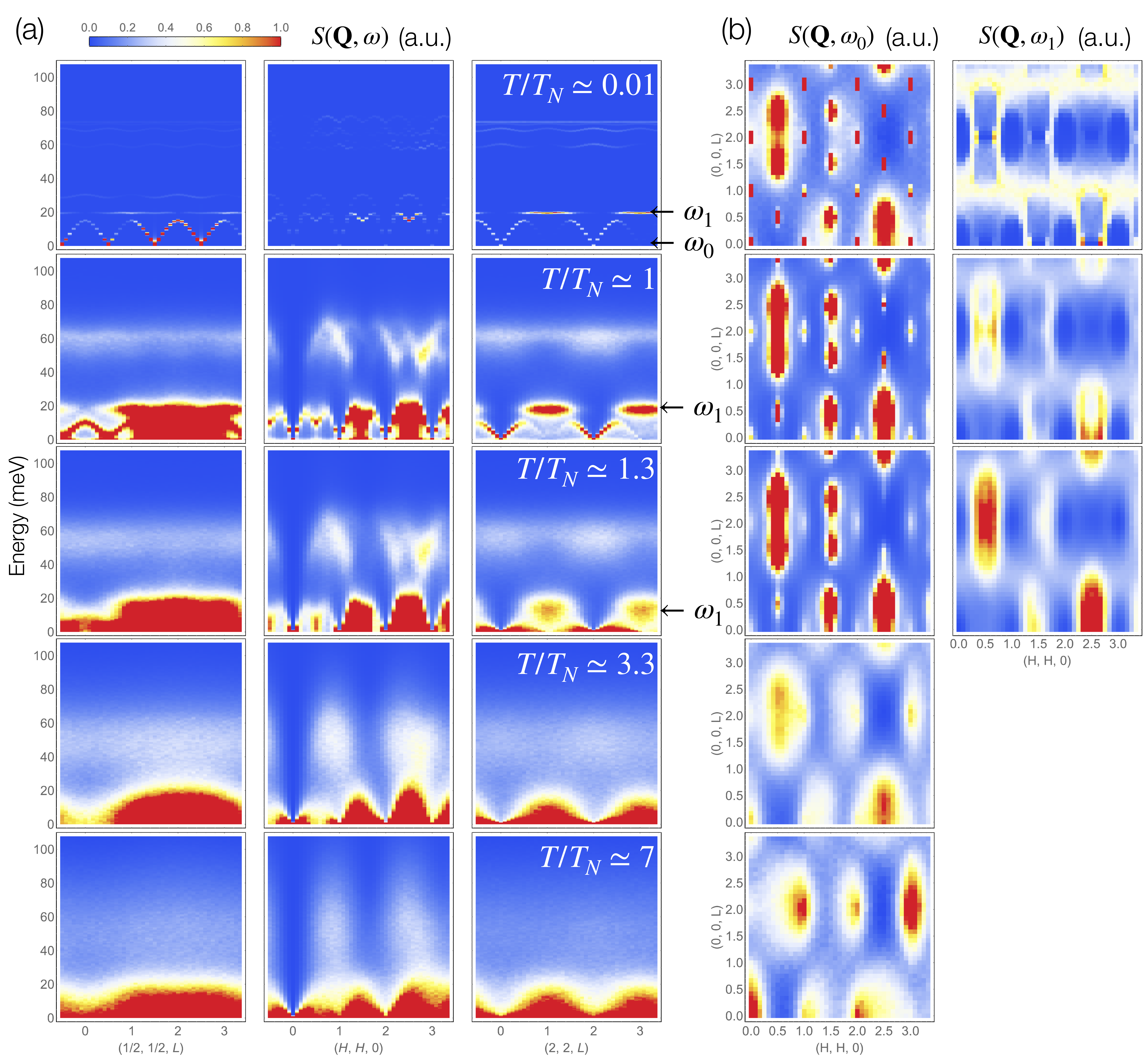}
	\caption{Calculated S(Q,$\omega$) by classical spin dynamics for different relative temperatures $T/T_N$ through (a) Q-cuts and (b) constant $\omega$-cuts, including zero energy $\omega_0$ and the energy corresponding to the nearly flat mode $\omega_1$ (indicated by the arrows). The intensity scale is the same for all the (a) maps and varies for the (b) maps in order to emphasize the main features.}
	\label{FigPlotSpinDynamicsCalc}
\end{figure*}

\begin{figure*}[h!]
	\includegraphics[width=1\columnwidth]{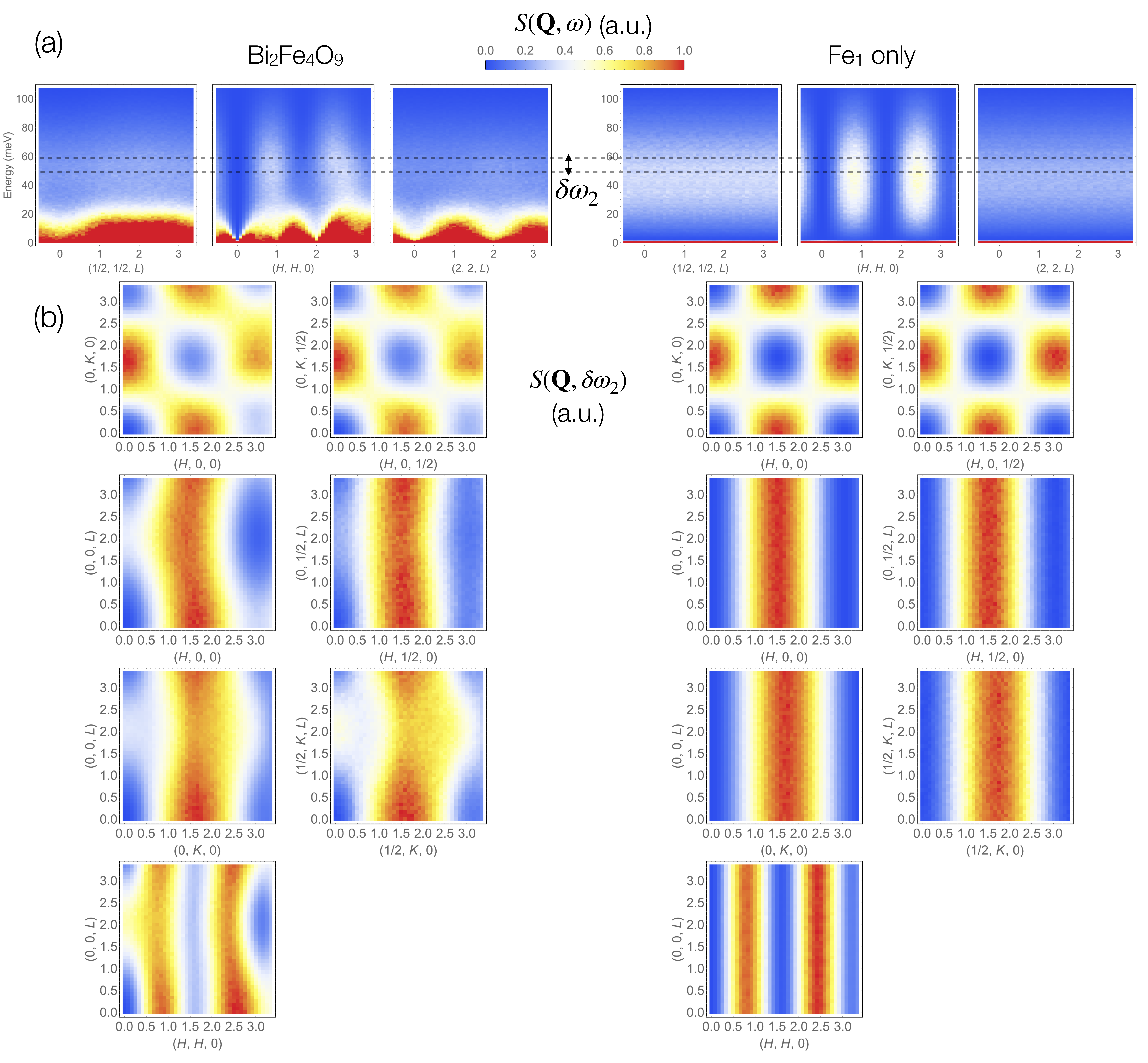}
	\caption{Comparison of the calculated S(Q,$\omega$) for Bi$_2$Fe$_4$O$_9$ using the set of exchange coupling obtained from our spin waves analysis with the one taking into account only the $Fe_1$ spins coupled via $J_4$, at $T/T_N\simeq7$ through (a) Q-cuts and (b) constant $\omega$-cuts over an energy range $\delta\omega_2$ (between the dotted lines). The intensity scale is the same for all the (a) maps and varies for the (b) maps in order to emphasize the main features.}
	\label{FigPlotFe1only}
\end{figure*}

We have computed the dynamical structure factor through classical spin dynamics simulations. More specifically, S(Q,$\omega$) was solved numerically by combining a Monte Carlo method to an integration of the equations of motion, therefore taking into account non-linear effects associated with thermal fluctuations (see Ref.  \cite{Taillefumier,Robert,Robert2} for technical details about the method). The calculations have been performed using the Hamiltonian determined by our spin wave analysis. Intensity maps of S(Q,$\omega$) are shown in Figure \ref{FigPlotSpinDynamicsCalc} for different relative temperatures. The evolution of the structure factor highlights strong spin pair correlations persisting even at temperature much higher than the N\'eel  temperature and several regimes in the paramagnetic state. Just above the N\'eel temperature, the Fe$_1$ and Fe$_2$ are correlated within the pentagonal planes but the correlations along the c axis through $J_2$ is lost. At higher temperatures, typically $7\times T_N$, only the Fe$_1$ dimers remain correlated forming a classical dimer state recalling the Shastry-Sutherland physics, but where the Fe$_1$ quasi-orthogonal dimers are separated by paramagnetic free $Fe_2$ spins. Finally, the system is expected to become completely uncorrelated at very high temperature. The dimer regime is identified from a comparison with a calculation taking into account only the Fe$_1$ spins, which are responsible for the high-energy features, as shown in Figure \ref{FigPlotFe1only}.

\clearpage

%

\end{document}